\documentstyle[twoside,12pt,epsf]{article}
\let\ssection=\section
\renewcommand{\section}{\setcounter{equation}{0}\ssection}

\def\head#1#2{
\markboth{}{}
\setcounter{page}{#1}
\setcounter{section}{0}
\begin{huge}
\begin{flushleft}
\noindent\hangindent\parindent
{#2}
\end{flushleft}

\end{huge}
\bigskip
}

\def\auaddr#1#2{
{\noindent\LARGE\em#1}
{\noindent #2}
\medskip}

\def\and{
{\LARGE\em \&}
\bigskip
}

\def\caption#1{{ \begin{quote} \rm #1 \end{quote} }}

\def\ben{\begin{equation}}
\def\een{\end{equation}}
\def\bea{\begin{eqnarray}}
\def\eea{\end{eqnarray}}
\input amssym.def
\input amssym.tex

\begin{document}
\thispagestyle{plain}
\def\tit{Supergravity vacua and solitons}
\def\auth{G W Gibbons}
\head{470}
{\tit}
\auaddr{\auth}{}
\markboth{G W Gibbons}{Supergravity vacua and solitons} \footnote{
Prepared for A Newton Institute Euroconference on Duality and 
Supersymmetric Theories, Cambridge, England, 7-18 Apr 1997.
published as   ``Cambridge 1997, Duality and supersymmetric theories''
eds P.West and D. Olive (1990) 267-296, Cambridge University Press}
   
\def\ve{\varepsilon}

\section{Introduction}

The enormous recent progress, detailed in other articles in this volume, 
on the non-perturbative
structure of string/M-theory
and its low energy approximation, quantum-field theory, 
has been based on the recognition of the importance of 
p-branes, i.e. of extended objects with p spatial dimensions.
Nowadays these can now be viewed in various ways. Historically however
they first appeared as classical solutions of the low energy limits,
SUSY Yang-Mills theory or the various Supergravity theories.
The purpose of these two lectures is to provide a pedagogic introduction
to  the properties of solitons in supergravity theories
and how one constructs the classical solutions. No claim is made for 
completeness, the subject is by now far too vast to survey the 
entire subject in just two lectures, and what follows is to some extent a 
rather personal account of the theory emphasising the features
which are peculiar to the gravitational context. Thus lecture
one is mainly concerned with exploring the question: what are the 
analogues of \lq solitons\rq in gravity theories? The second
is concerned with  finding p-brane solutions. 
An important subsidiary technical theme is the use of sigma models
or harmonic maps to solve Einstein's equations.
Most, but not all, of the material is about four spacetime dimensions,
partly because this is the best understood case and the most 
extensively studied and partly because, at least my, intuition
is strongest in that dimension.

My own interest in this subject began in the early days of 
supergravity theories with the realization that since perturbatively
such theories describe just a system of interacting massless
particles: the graviton $g_{\mu \nu} $
$N$-gravitini $\psi^i_\mu$, , $n_v$ abelian vectors $A^A _\mu$,
$n_{1\over 2}$ spin half fermions $\lambda $
and $n_s$ scalars ${\bar \phi}^a$
, then unless non-perturbative effects
come into play they can have little connection with physics.
It soon became apparent to me that the non-perturbative structure
must involve gravity in an essential way, and influenced by some ideas
of Hajicek, which he mainly applied to gravity coupled
to non-abelian gauge theory, I was led to propose extreme black holes
as the appropriate analogue of solitons, what we now call BPS states.
With this in mind I embarked in the early 1980's on a series of 
investigations aimed at uncovering the essential 
features of the soliton concept
in gravity. What follows is largely a synopsis of those ideas
updated to take into account recent developments. 
It has been both  gratifying and a little surprizing to see 
how well they have remained relevant over the intervening years.

\section{The soliton concept in SUGRA theories}

\subsection{SUGRA in 4-dimensions}

In what follows we shall concentrate on the theory in four spacetime
dimensions but much of what follows goes through
with appropriate modifications in higher dimensions.
Some examples of this will be given {\it  en passant}.

\subsubsection{The Lagrangians}
If spacetime $\{ {\cal M}, g_{\mu \nu} \}
$ has  four dimensions the lagrangian $L=L_F+L_B$ of an
un-gauged  
supergravity theory with $N$ supersymmetries
has a fermionic $L_F$ and bosonic $L_B$ part.
The latter is
given by
\ben
L_B=  { 1\over 16 \pi } R-{ 1 \over 8 \pi } 
{\bar G}_{ab}g^{\mu \nu}\partial_\mu {\bar \phi}^a \partial_\nu {\bar \phi} ^b - { 1\over 16 \pi} \mu({\bar \phi})_{AB} F^A_{\mu \nu} {F^B}^{\mu \nu}
- { 1 \over 16 \pi } \nu({\bar \phi})_{AB} F^A_{\mu \nu} \star {F^B}^{\mu \nu}
\een
where $R$ is the Ricci scalar of the spacetime metric $g_{\mu \nu}$,
$\star$ is the Hodge dual. As is usual when combining
gravity and electromagnetism, I am  using Guassian or 
\lq un-rationalized\rq  units so as to avoid extraneous factors of $4 \pi$ is the formulae. We have also set Newton's constant $G=1$. 
Occasionally I will, without further comment, re-instate it.

I have adopted the 
spacetime signature $(-1,+1,+1,+1)$. The particular 
convenience of that signature choice is that because 
the Clifford algebra ${\rm Cliff}_{\Bbb R} (3,1) \equiv {\Bbb R}(4)$
the algebra of four by four real matrices, {\sl everything} 
in the purely 
classical theory, including all spinors and spinor lagrangians, may be taken to be real. The same is of course true of classical M-theory
since ${\rm Cliff}_{\Bbb R} (10,1) \equiv {\Bbb R}(32)$
and continues to hold if we descend to ten spacetime dimensions.
It seems that it is only when we pass to the quantum theory
that we need to introduce complex numbers. When dealing with
Grassmann algebra valued spinors in classical supersymmetric
theories one need only consider 
algebra's over the field of real numbers.
However the consistent adoption of this
point of view  entails some minor
changes of the conventions, concerning for example \lq conjugation \rq
which are customary in the subject
because $i$'s never appear in any formulae. 
My own view is the customary conventions
introduce into the theory extraneous and unnatural 
elements which disguise the underlying mathematical simplicity.
Fortunately, perhaps, the  explicit details
of the necessary changes in conventions 
will not be needed in what follows.

\subsubsection{ The scalar manifold} 

The scalar fields ${\bar \phi} ^a$, $a=1,\dots,n_s$ take values in 
some target manifold
$M_{\bar \phi} $ with 
metric ${\bar G}_{ab}$ which is typically a non-compact
{\it symmetric space} ,
$M_{\bar \phi}= {\bar G} /{\bar H}
$ where ${\bar H} $ is the maximum compact subgroup of ${\bar G}$
and the Lie-algebra ${\bar \frak{g}} = {\bar \frak{h}}\oplus {\bar
\frak{k}} $ admits the 
involution $({\bar \frak{h}} , {\bar \frak{k}} )\rightarrow ({\bar \frak{h}}, -{ \bar \frak{k}} )$.
As a consequence 
\begin{itemize}
\item topologically $M_{\bar \phi}
$ is trivial, $M_{\bar \phi} \equiv 
{\Bbb R}^{n_s}$.
\item 
The sectional curvatures $K(e_1,e_2)=R_{abcd}e^a_1e^b_2e^c_1e^d_2$ 
are non-positive. 
\end{itemize}

These conditions play an important role in the theory of harmonic
maps. For example a well known theorem states that
the only regular
harmonic maps
from  a compact Riemannian manifold of positive Ricci curvature
to a compact Riemannian manifold of negative sectional curvature
are trivial. It is not difficult to adapt the proof 
of this theorem to rule out static
asymptotically flat
Skyrmion
type solutions of the purely gravity-scalar sector of the theory.
As a consequence we can anticipate that there are few
non-perturbative features of the purely scalar sector.
This situation changes when one  passes to the gauged SUGRA
theories. In that case there is a potential function
for the scalars
and various domain wall configurations exist.
I shall not pursue this aspect of the theory further here.

\subsubsection {Duality}

The abelian vector fields $A^A_\mu$
, $A=1,\dots, n_v$ transform under a representation of $G$ and and
may be thought of as sections of the pull back under the map ${\bar \phi}
: {\cal M}  \rightarrow M_{\bar \phi}$ of a vector bundle over 
$M_{\bar \phi}$  tensored with spactime one-forms. 
In fact the structure is somewhat richer.  As explained by Zumino
elsewhere in this volume, one may define an 
electromagnetic induction 2-form by:
\ben
 G_A^{\mu \nu} = -8 \pi {\partial L_B \over \partial F^A_{\mu \nu}}.
\een
The pair $(F^A, \star G_A)$ also carries a  representation of 
$G$ and in fact may be 
considered as the pull-back of an $Sp(2n_v, {\Bbb R})$ bundle
over $M_{\bar \phi} $ tensored with spacetime two-forms. 

To see what is going on more
explicitly it is helpful to cast the vector lagrangian in
non-covariant form.  In an orthonormal frame it becomes,
up to a factor of $4 \pi$
\ben
{1 \over 2} \mu_{AB}\bigl (  {\bf E}^A \cdot{\bf E}^B -{\bf B}^A \cdot
{\bf B}^B \bigr ) 
+ \nu _{AB} {\bf E}^A \cdot {\bf B}^B. 
\een
We define 
\ben
{\bf D}_A= \mu_{AB} {\bf E}^B + \nu_{AB} {\bf B}^B
\een
and
\ben
{\bf H}_A= \mu_{AB} {\bf B}^B - \nu_{AB} {\bf E}^B.
\een
Thus
\ben
\pmatrix{ {\bf H}_A \cr {\bf E}^A \cr}
=\pmatrix {\mu + \nu \mu ^{-1} \nu  & - \nu \mu ^{-1} \cr
 - \mu ^{-1} \nu & \mu ^-{-1}  \cr } 
\pmatrix{ {\bf B}^A \cr {\bf D}_A \cr}.
\een
We define the matrix
\ben {\cal M} ({\bar \phi } ) = \pmatrix {\mu + \nu \mu ^{-1} \nu  & - \nu \mu ^{-1} \cr
 - \mu ^{-1} \nu & \mu ^-{-1}  \cr },
\een
and finds that
\ben
{\rm det} {\cal M} ({\bar \phi } )=1.
\een

The local density of energy due to the vectors is 
\ben
{ 1\over 8 \pi} \bigl ({\bf H}_A \cdot {\bf B}^A+ {\bf D}_A \cdot {\bf E}^A  \bigl ). 
\een

The energy density and the equations of motion,  will be  invariant under
the action of $ S\in SL(2, {\Bbb R})$
\ben
\pmatrix{ {\bf B} \cr {\bf D} \cr} \rightarrow S \pmatrix{ {\bf B} \cr {\bf D} \cr}
\een
provided we can find an action 
of $SL(2, {\Bbb R}) $
on the scalar manifold $ S: {M} _{\bar \phi} \rightarrow  { M} _{\bar \phi}$  which leaves the metric ${\bar G} _{ab}$ invariant
and under which the matrix ${\cal M } 9 {\bar \phi }) $ pulls back as
\ben
{\cal M} \rightarrow (S^t) ^{-1}   {\cal M } S^{-1}.
\een
As explained by Zumino this happy situation 
does inded prevail in the SUGRA
models under present consideration. We note {\it en passant}
that electric-magnetic duality transformations 
of this type may extended to  the {\sl non-linear } electrodynamic
theories of Born-Infeld type which one encounters 
in the world volume actions of Dirichlet 3-branes.

The set-up described above may be elaborated
somewhat . For general $N=2$ models,
including supermatter, $M_{\bar \phi}$ is a K\"ahler manifold subject to
\lq Special Geometry\rq. The properties
described in the present article do not require this extra structure
and it will not be necessary to understand precisely what special geometry
is in the sequel.

\subsubsection{Scaling symmetry}

As mentioned above, perturbatively
theories of this kind describe massless particles.
In fact, since the vectors are {\it abelian}, even more is true. The theory
admits a global {\it scale-invariance}:
\ben
(g_{\mu\nu}, F^A_{\mu\nu}, {\bar \phi} ^a) \rightarrow (\lambda^2 g_{\mu\nu}, \lambda F^A_{\mu\nu}, {\bar \phi}^a)
\een
with $\lambda \in {\Bbb R}_+$ which takes classical
solutions to classical solutions
and takes the lagrangian density $\sqrt{-g} L_B \rightarrow \lambda ^{2} 
\sqrt{-g}
L_B$.

It is an easy exercise to check that our general
lagrangian has an energy momentum tensor $T^\mu _\nu$ which
satisfies the {\it dominant energy condition},
i.e $T^\mu _\nu p ^\nu$ lies in or on the future light cone
for all vectors $p^\nu$ which themselves lie
in or on the future light cone. Thus these theories
have good  local stability and causality properties.  

Given the scaling symmetry and the positivity of energy
one may construct  arguments analogous to the well known
theorem of Derrrick in flat space to show that
there are no non-trivial everywhere static or stationary
solutions of the field equations. This type of theorem,
are often referred to as Lichnerowicz theorems, 
though they go back to Serini, Einstein and Pauli.
They may be summarized by the slogan: {\bf  No solitons without
horizons} .

\subsubsection {Examples}

At this point some examples of ${\bar G} /{\bar H}$ are in order.
\begin{itemize}

\item $N=1$. We only have a graviton,
$n_s=0=n_v$, 
and thus bosonically this is just ordinary General Relativity.

\item $N=2$, $U(1)/U(1)$, there are no scalars, $n_s=0$ and one vector, $n_v=1$.  This is just
Einstein-Maxwell theory

\item $N=4$, $SO(2,1) /SO(2) \times SO(6) /SO(6)$.
We have two scalars and six vectors.

\item $N=4$ plus supermatter, $SO(2,1) /SO(2) \times 
SO(6,6) /SO(6)\times SO(6)$. This is what you get if you
dimensionally reduce $N=1$ supergravity from ten-dimensions.

\item The reduction of the Heterotic theory in 10-dimensions,
$SO(2,1) /SO(2) \times 
SO(22,6) /SO(22)\times SO(6)$

\item $ N=8$, $E_{7(7)} /SU(8)$.

\end{itemize}

Note that, except for $N=8$, $G=S\times T$ where $S=SL(2,{\Bbb R})$ is 
what is now called the S-duality group and $T$ is now called
the T-duality group. In the case of $N=8$ both $S$ and $T$ are contained
in a Hull and Townsend's single $U$-duality group.

\subsection{Vacua}

Presumably the minimum requirement of a classical {\it vacuum} or 
{\it ground state} is that it be a homogeneous spacetime 
${\cal M} = {\cal G} / {\cal H}$
with constant
scalars , $\partial_\mu {\bar \phi} ^a=0$ and covariantly constant Maxwell fields
$\nabla _\mu F^A_{\nu \rho}=0$. Actually one sometimes wishes to consider a 
\lq linear dilaton\rq but we shall not consider that possibility here.
The ground state will this be labelled in part by a point 
$p\in M_{\bar \phi}$.
Thus one may consider $M_{\bar \phi}$, or possibly some sub-manifold
of it, as parameterizing a {\it moduli space}
of vacua. We usually demand that the ground state be classically
stable,
at least against small disturbances and this typically translates into
the requirement that it have least \lq energy \rq  among all 
nearby solutions with the same asymptotics. 
Unfortunately there is no space here to enter in detail
into a general discussion
of the how energy is defined in general relativity
using an appropriate
globally timelike Killing vector field $K^\mu$. Later we shall
outline an approach based on supersymmetry and Bogomol'nyi bounds.
Suffice it to emphasise at present two important general principles.
\begin{itemize}
\item The dominant energy
condition plays an essential role in establishing the positivity.
\item Stability cannot be guaranteed
if there is no globally timelike Killing field $K^\mu$. 
\end{itemize}
Thus if spacetime
${\cal M}$ admits non-extreme Killing Horizons across which 
a Killing field $K^\mu$ switches from being 
timelike to being spacelike the local energy momentum vector  
$T_{\mu _\nu} K^\nu$  {\sl relative to
$K^\mu$ } can become spacelike.  
This fact is behind the quantum instability
of non-extreme Killing horizons due to Hawking radiation. 
In the context of homogeneous spacetimes this means that while
globally static Anti-de-Sitter spacetime $AdS_n=SO(n-1,2)/SO(n-1,1)$
has positive energy properties,
and can  be expected to be stable,  de-Sitter  
spacetime $dS_n =SO(n+1,1)/SO(n-1,1)$ with its 
cosmological horizons does not admit a global
definition of positive energy.

For the class of lagrangians we are considering
in four spacetime dimensions,
there  are two important ground states.
One is familiar Minkowski spacetime ${\Bbb E}^{3,1}= E(3,1)/SO(3,1)$,
for which $F^A_{\mu \nu}=0$ and ${\bar \phi} ^a$ is arbitrary.
The other, probably less familiar, is Bertotti-Robinson spacetime
$AdS_2 \times S^2 = SO(1,2)/SO(1,1) \times SO(3)/SO(2)$.
This represents a \lq compactification \rq from four to two spacetime
dimensions on the two-sphere $S^2$.  One has
\ben
F^A= { p^A \over A} \eta_{S^2}
\een
and
\ben
\star G_A= {q_A \over A}\eta_{S^2}
\een
where $ \eta_{S^2}$ is the volume form on the $S^2$ factor,
$A $ is its  area, $(p^A,q_A)$ are constant
magnetic and electric charges and the scalar field ${\bar \phi} ^a$ is \lq
frozen \rq at a value ${\bar \phi}^{\rm frozen}$
which extremizes a certain potential
$V({\bar \phi}, p,q)$ which may be read off from
the lagrangian. The potential is given by
\ben
V({\bar \phi}, p,q)= \pmatrix { p & q \cr}  {\cal M} ( {\bar \phi})
\pmatrix { p \cr q \cr }   
\een
and is invariant under S-duality which acts on the charges as 
\ben
\pmatrix { p \cr q \cr } \rightarrow S \pmatrix { p \cr q \cr }.
\een
The frozen values of ${\bar \phi}$ are given by
\ben
{\partial V({\bar \phi} , p,q) \over {\bar \phi} } \Bigr |_{{\bar \phi}
={\bar \phi} ^{\rm frozen}} =0.
\een
In fact
\ben
A= 4 \pi  V({\bar \phi} ^{\rm frozen}, p,q).
\een
Thus if $V({\bar \phi}, p,q) $ has a unique minimum
value then the vacuum solution is specified entirely by giving the charges
$(p,q)$.

As we shall see, it turns out that almost all of the black hole
properties of the theory
are determined entirely by the function $V({\bar \phi}, p,q) $.
In particular SUGRA theories, such as $N=2$ theories
based on Special Geometry, one  may know some special facts 
about $V({\bar \phi}, p,q)$ and one may read off much
of the nature of the BPS configurations directly,
without a detailed investigation of particular
spacetime metrics.

\subsection {Supersymmetry and Killing Spinors}

The supersymmetry transformations  take the  schematic form:
\ben \delta_\epsilon B=F
\een
\ben \delta_\epsilon F=B
\een
where $(B,F)$ are bosonic and fermionic fields respectively
and $\epsilon=\epsilon^i$, $i=1,\dots, N$ is an N-tuplet of spinor fields. 
A purely bosonic background $(B,0)$ is said to admit a Majorana Killing spinor
field $\epsilon$, or to admit SUSY, if
\ben
\delta_\epsilon (B,0)=0.
\een
Because the Killing spinor condition is linear in  spinor fields,
we may take the Grassmann algebra valued spinor fields
$\epsilon$ to be  commuting spacetime dependent spinor fields 
(which by an abuse of notation we also call $\epsilon$)
multiplied 
by a constant Grassmann coefficient. 
The Killing spinor condition reduces to $\delta_\epsilon F=0$,
or in components:
\ben
\delta_\epsilon \psi ^i\mu = ({\hat \nabla} _\mu \epsilon)^i
\een
\ben
\delta_\epsilon \lambda = M \epsilon
\een
where $\psi ^i_\mu$ are the $N$ gravitini and $\lambda$
are the spin half fields. The operator ${\hat \nabla} _\mu= \nabla_\mu + E_\mu$
where $\nabla_\mu$ is the Levi-Civita covariant derivative acting on spinors
and $E_\mu$ are endomorphisms depending on the bosonic field and whose
precise form depends upon the particular SUGRA theory under
consideration.

It follows from the supersymmetry algebra that
\ben
K_{\epsilon}= {\bar \epsilon} \gamma ^\mu \epsilon
\een
is a Killing vector field which necessarily lies in or on the future 
light cone.
The solution is said to admit maximal SUSY if the dimension
of the space of Killing spinors is $N$. If it is less, we speak of
have BPS
states with ${N \over k}$ SUSY. From the point of view of 
SUSY representation theory BPS solutions correspond to short
multiplets.

Typically  {\sl both } 
Minkowski spacetime ${\Bbb E}^{3,2}$ {\sl and} Bertotti-Robinson spacetime
$AdS_s \times S^2$ admit maximal SUSY. 
In gauged supergravity one has either a negative cosmological constant
or a negative potential for the scalars. It then turns
out  that anti-de-Sitter spacetime
is a ground state with maximal SUSY. Because
it does not admit an everywhere causal 
Killing field, de-Sitter space, even if it were a solution, 
could   never  admit SUSY.

\subsubsection { Remark on Five dimensions} 

For the purposes
of discussing black hole entropy it is often
simpler to 
treat five-dimensional supergravity theories.
I will not discuss them here in detail. I will simply remark
that 
much of the present 
theory goes through.
After dualization the bosonic lagrangian is 

\ben
L_B=  { 1\over 16 \pi} R-{ 1\over 8 \pi} G_{ab}g^{\mu \nu}
\partial_\mu \phi^a \partial_\nu \phi^b - { 1\over 16 \pi} \mu(\phi)_{AB} F^A_{\mu \nu} {F^B}^{\mu \nu}
+ {\rm Chern-Simons \thinspace term}.
\een
and $AdS_2 \times S^2$ is replaced by $AdS_2 \times S^3$. For example if
$N=8$, $G/H= E_{6,(+6)}/Usp(8)$.

\subsection {pp-waves}

The perturbative, massless, states of the theory correspond
to wave solutions. For example 
\ben
ds^2 = -dt^2 + dz^2 + dx^2 + dy^2 + K(t-z, x,y) ( dt-dz )^2
\een
with $K(t-z,x,y)$ being a harmonic function of $(x,y)$
\ben
(\partial _x^2 +\partial_y^2) K=0 
\een
with arbitrary dependence on on $t-z$, is a vacuum solution representing
a classical gravitational wave propagating in the positive $z$ direction.
Such solutions are called \lq pp-waves \rq . As one would expect from basic
supersymmetry representation theory they admit $N={ 1\over 2}$ SUSY.
If 
\ben
\bigl (\gamma ^t - \gamma^z \bigr ) \epsilon=0
\een
then $\epsilon$ is a Killing spinor and by suitably scaling it it we
have
\ben
{\bar \epsilon} \gamma ^ \mu \epsilon = K^\mu
\een
where
\ben
K= {\partial \over \partial t} +{\partial \over \partial z}  
\een
is the lightlike Killing field of the metric.

This example admits an obvious generalization to arbitrary
spacetime dimension $n$. One simply replaces the two transverse
coordinates $(x,y)$ by $n-2$ transverse coordinates 
$(x^1, \dots x^{n-2})$.
If $K$ is taken to be independent of $t-z$ and one sets $H=1+K$ one may
dimensionally reduce to $n-1$ spacetime dimensions:
\ben
ds^2= H ( dz-{dt \over H} )^2 - 
{ 1 \over H^{1 \over n-3} }
 \Bigl (-{ dt^2 \over H^{n-3}} + H^{ 1 \over n-3} dx_{n-2}^2 \Bigr ).
\een
This gives a Kaluza-Klein 0-brane, with $A= { dt \over 2H}$.
For example if $n=5$ we get a so-called $a=\sqrt{3}$ 
extreme black hole which is S-dual to a the Kaluza-Klein monopole
based on the Taub-NUT metric. If $n=11$ we get the Dirichlet zero-brane of 
ten-dimensional type IIA theory.

\subsection{ Asymptotically flat solutions}

We suppose that $g_{00} \sim -1+ { 2GM \over r}$ and $g_{ij} \sim (1+{2GM \over r} ) \delta_{ij}$ where $M$ is the ADM mass. 
The scalar field ${\bar\phi}
\sim {\bar\phi}_\infty + {\Sigma ^a \over r}$ where the the scalar charge
$\Sigma ^a \in TM_{{\bar \phi} _\infty}$. 
Because the scalar manifold $M_{\bar \phi}$ is topologically trivial
and because its sectional curvature is non-positive
it is not difficult to prove, using standard techniques from the 
theory of harmonic maps, that there are no non-singular
solutions without horizons  
with vanishing vector fields.  Thus there are no analogues
of Skyrmions.

We define the total electric $q_A$ and magnetic $p^A$ by the usual
2-surface integrals at infinity, 
\ben
p^A= { 1\over 4 \pi}  \int_{S^2_\infty} F^A
\een
and
\ben
q_A= { 1 \over 4 \pi} \int_{S^2_\infty} \star G_A.
\een

We now arrive at some absolutely crucial points. 
\begin{itemize}
\item The
fundamental fields carry neither electric nor electric charges.
Thus perturbative states cannot carry them. 
In fact, by Maxwell's equations, 
a solution can only carry non-vanishing charges
if it is in some way singular or topologically non-trivial or both.
In fact we encounter here the phenomenon of 
\lq charge without charge \rq due to \lq lines of force being trapped
in the topology of space \rq which formed such central point of 
Misner and Wheeler's \lq Geometrodynamics\rq. In the context of 
string theory and Polchinkski's non-perturbative
Dirichlet-branes, 
some of the vector fields have their origin  
in the   Ramond$\otimes$Ramond sector and we have \lq Ramond$\otimes$Ramond
charge without charge \rq.

\item Because the fields are abelian
and also because of the classical scale-invariance
there is no possibility of a classical quantization
of the charges. Quantization can only be achieved by going outside
the framework of classical supergravity theory, for example
by coupling to so-called \lq fundamental\rq branes
and applying a Dirac type argument 
or by applying Saha's well known argument for the angular momentum
about the line of centre's joining
an electric and magnetic charge. One then discovers that if
the electric charges belong to some lattice $q_A \in \Lambda$ 
then the magnetic charges belong to the reciprocal lattice $p^A \in \Lambda ^\star$. That is
\ben
{ 2 \over \hbar } { p^A q_A } \in {\Bbb Z}.
\een
\end{itemize}

\subsection{Black Holes}
In the case of  $N=1$ one might be tempted
to think that the only  static solution with
a regular event horizon, the Scharzschild black hole
should be considered as some sort of soliton. However there are a number of
reasons why this is not correct.
\begin{itemize}

\item Although the solution is 
regular outside its event horizon,
inside it contains a spacetime singularity. This may not be fatal
if Cosmic Censorship holds. 
In that case the spacetime out side the event horizon
would be regular and predictable. In fact it is widely believed
to be classically stable.
\item However the mass $M$ is arbitrary and is
not fixed by any quantization condition. 
Moreover classically the black hole can absorb gravitons 
and gravitini leading to a mass increase. In fact classically the 
area $A$ of the event horizon can 
never decrease. This irreversible behaviour is quite unlike what one expects of a classical soliton.
\item Quantum mechanically, because of the Hawking effect
the Schwarzschild black hole is definitely unstable. 
The same is true of the Kerr solution. It does not seem reasonable therefore
to expect that in the full quantum gravity theory
one may associate with them a stable non-perturbative
state in the quantum mechanical Hilbert space.
In fact because of their thermal nature it is much more likely
that the classical solutions should be associated with a density matrices
representing a black hole in thermal equilibrium with its evaporation
products.
\item From the point of view of SUSY it is clear that
the Schwarzschild and Kerr solutions do not correspond to BPS states since they
do not admit any Killing spinors. For $N=1$ the endomorphism $E_\mu$
vanishes and such killing spinors would have to be covariantly constant,
as would the Killing vector constructed from them
This is impossible if the solution is not to be flat.

\end{itemize}
Later we shall see that SUSY is incompatible with a non-extreme
Killing horizon.

\subsubsection{ Reissner-Nordstrom} 
If $N=2$ the candidate static
solitons would be Reissner Nordstrom black holes.
If the singularity is not to be naked we must have

\ben
M \ge |Z|,  
\een
where
\ben
|Z|^2={q^2 + p^2 \over  G}. 
\een
Note that in this section I am re-instating Newton's constant $G$.
The Hawking temperature is 
\ben
T= { 1\over 4 \pi G} { \sqrt{ M^2-|Z|^2} \over (M+ \sqrt{ M^2-|Z|^2} )^2 }.
\een
If $M >|Z|$ the temperature is non-zero and
the solution is unstable against Hawking evaporation
of gravitons, photons and gravitini. Moreover it cannot be a BPS state
since the Killing vector which is timelike near infinity 
becomes spacelike inside the horizon.

Only  in the extreme case $M=|Z|$ for which $T=0$
is $\partial \over \partial t$ never
spacelike. Moreover in that case there are multi-black hole solutions,
the so-called Majumdar-Papapetrou solutions:
\ben
ds^2= -H^{-2} dt^2 + H^2 d{\bf x}^2,
\een
with, up to an electric-magnetic  duality rotation,
\ben
F=dt\wedge d ( { 1\over H}),
\een
where $H$ is an arbitrary  harmonic function on ${\Bbb E}^3$.

It is an easy exercise to verify that 
the entire Majumdar-Papapetrou family of solutions admits a Killing spinor
whose associated Killing vector is $\partial \over \partial t$.
In fact they are the only static solutions of Einstein-Maxwell
theory admitting a Killing spinor. Thus the Majumdar-Papapetrou
solutions correspond to BPS states.

One further property of the extreme holes,
called {\it vacuum interpolation} , should be noted. This
is while near infinity the solution tends to
the flat maximally supersymmetric ground state, 
Minkowski spacetime ${\Bbb E}^{3,1}$,
near the horizon the metric tends to the other maximally
supersymmetric ground state, Bertotti-Robinson spacetime
$AdS_2 \times S^2$. Thus, as is the case for  many solitons,
the solution spatially interpolates between different vacua or 
ground states of the theory.

The Bekenstein-Hawking entropy $S$ of a 
general charged black hole is given by
\ben
S= { A\over 4G} = \pi G(M+ \sqrt{M^2-|Z|^2} )^2.
\een
For fixed $|Z|$ this is least in the extreme case when
it attains
\ben{ \pi } (q^2 + p^2),
\een
which is {\sl independent} of Newton's constant $G$ and depends {\sl only}
on the quantized charges $(p,q)$.

\subsubsection { Black holes and frozen moduli}

If we now pass to the case when more than one vector and some 
scalars are present we find that in general 
that the scalars will vary with position. There is thus
in a sense 
\lq scalar hair \rq.  However   spatial dependence of the 
 scalar fields and non-vanishing
scalar charges $\Sigma ^a$ are only
present by virtue of the fact that the source term
$ {\partial V \over {\bar \phi }} ({\bar \phi},p,q)$
in the scalar equations of motion. If
it happens that ${\bar \phi}_\infty = {\bar \phi } ^{\rm frozen}$
however then the charges vanish and the scalars are constant.
The moduli are then said to be frozen. The geometry of the
black holes is then the same as the Reissner Nordstrom case
with
\ben
Z^2= V({\bar \phi} ^{\rm frozen}, p,q) = |Z |^2(p,q).
\een

In general the moduli will not be frozen 
and for instance the value of the scalars on the horizon 
${\bar \phi}^{\rm horizon}$ will be different from
its value ${\bar \phi}_\infty$ at infinity. For extreme static black holes
however regularity of the horizon 
demands that 
\ben
{\bar \phi}^{\rm horizon}= {\bar \phi} ^{\rm frozen}. 
\een

As a consequence we have the important general fact,
 that
the Bekenstein-Hawking entropy of extreme holes is always independent
of the moduli at infinity and depends only on the
quantized charges $(p,q)$. That is 
\ben
S_{\rm extreme} = \pi V({\bar \phi},p,q)= \pi |Z|^2(p,q).  
\een

As explained in other lectures the Bekenstein-Hawking
entropy of extreme holes can be obtained by D-brane 
calculations at weak coupling. It is vital for the
consistency of this picture that  $S_{\rm extreme}$ really is independent
of the moduli ${\bar \phi}_\infty$ which label the vacua. 
A striking consequence of this is that the entropy
of any initial data set with given $(p,q)$ should never be less
than $\pi | Z(p,q)|^2$ andits mass $M$ should never be less than
$|Z(p,q)|$.

\subsubsection{The First Law of Thermodynamics and the Smarr-Virial Theorem}

For time stationary fields we may define electrostatic potentials
$\psi^A$ and magnetostatic potentials $\chi_A$ by
\ben
F^A_{0i}= \partial _i \psi^A
\een
and
\ben
{G_A} _{0i}= \partial _i \chi_A.
\een
The first law of classical black hole mechanics needs a modification
if we consider variations of the moduli ${\bar \phi}_\infty$.
It becomes
\ben
dM= T dS + \psi^A dq_A + \chi_A dp^A - \Sigma ^a 
{\bar G } _{ab} ( {\bar \phi}_{\infty} )
d {\bar \phi} ^b.
\een
The last term is the new one. Note that, at the risk of causing confusion
but in the interest of leaving the formulae comparatively uncluttered
I have not explicity distinguished between the potential functions
 $\chi_A, \psi ^A$ and their values at the horizon.

It  follows using the scaling invariance
that the mass is  given by the Smarr formula 
\ben
M=2TS + \psi^Aq_A + \chi_A  p^A.
\een
Thus in the present 
circumstances, the Smarr formula
is equivalent to the first law as a consequence of scaling symmetry.
In other words we may regard the Smarr relation as a virial
type theorem. This formula allows a simple derivation of
the \lq No solitons without horizons \rq result.
If there is no horizon then $S=0$ and by Gauss's theorem $q_A=0=p^A$.
It follows that the mass $M=0$ and hence by the 
positive mass theorem the solution must be flat.

\subsection {Bogomol'nyi Bounds}
We shall now give a brief indication
of how one identifies
the central charges and establishes  Bogomol'nyi
Bounds. Let $\epsilon^i _\infty$ be constant spinors at infinity.
In what follows we shall sometimes omit writing out a summation of $i$
explicitly. 
The supercharges $Q^i$ are defined by
\ben
{\bar \epsilon _ \infty } Q
 = { 1 \over 4 \pi G} \int _ { S^2 _ \infty }
{ 1\over 2} 
{\bar \epsilon _\infty }  
\gamma ^{\mu \nu \lambda } 
\psi _\lambda  d \Sigma _{\mu \nu}. 
\een
The Nester two-form $N^{\mu \nu}$
associated to a spinor field $\epsilon$
is defined by
\ben
N^{\mu \nu}={\bar \epsilon  } \gamma ^{\mu \nu \lambda }
{\hat \nabla}  _ \lambda \epsilon.
\een
Under a SUSY variation we have 
\ben
{\bar {\epsilon _\infty }} \delta _{\epsilon} Q
= { 1 \over 4 \pi G} \int _ { S^2 _\infty}
{ 1\over 2} N^{\mu \nu}  d \Sigma _{\mu \nu}.
\een
Stokes's theorem gives
\ben
{\bar \epsilon _\infty } \delta _{\epsilon} Q
= { 1 \over 4 \pi G} \int _ \Sigma   \nabla _\mu N^{\mu \nu} d \Sigma _\nu,
\label{positive}\een
where $\Sigma$ is a suitable spacelike surface whose boundary
at infinity is $S^2_\infty$ and whose inner boundary either vanishes
or is such that by virtue of suitable boundary conditions
one may ignore its contribution.

Now by the supergravity equations 
of motion one finds
\ben
\nabla _\mu N^{\mu \nu} 
= {\bar { {\hat \nabla} _\mu \epsilon}} \gamma ^{\mu \nu \lambda } 
{\hat \nabla } \epsilon + {\bar {M \epsilon }} \gamma ^\nu M \epsilon
\een
Now restricted to $\Sigma$, $ {\bar M \epsilon} \gamma ^0 M \epsilon \ge
0 $
and
\ben
{\bar { {\hat \nabla} _\mu \epsilon} } \gamma ^ { \mu 0 \lambda } 
{\hat \nabla } \epsilon= | {\hat  \nabla} _a \epsilon |^2 -
| \gamma ^ a {\hat \nabla }_a \epsilon |^2,
\een
where the derivative ${\hat \nabla }_a$ is tangent
to  $\Sigma$. So far $\epsilon $ has been arbitrary. We pick it such
that
\begin{itemize}
\item $ \gamma ^a {\hat \nabla }_a \epsilon =0$
\item $\epsilon \rightarrow \epsilon _\infty $ at infinity.
\end{itemize}
We also choose $\epsilon$ such that the inner boundary
terms, such  as might arise at an horizon, vanish.
It is not obvious but it is in fact true that this can be done.
It follows that the right hand side of \ref{positive}
is non-negative and vanishes if and only if everywhere
on $\Sigma$ 
\ben
M \epsilon=0
\een
and
\ben
{\hat \nabla }_a \epsilon =0.
\een
Because $\Sigma$ is arbitrary we may deduce that in fact
 the right hand side of \ref{positive}
is non-negative and vanishes if and only if everywhere
in spacetime
\ben
M \epsilon=0
\een
and
\ben
{\hat \nabla }_\mu \epsilon =0.
\een
This means that $\epsilon$ must be  a Killing Spinor.

Now the left hand side of \ref{positive}
may be shown to be
\ben
{\bar { \epsilon _\infty }} \gamma ^\mu P_\mu  \epsilon _\infty
+ {\bar{ \epsilon _\infty } } \bigl ( U_{ij} + \gamma _5 V_{ij} \bigr ) \epsilon _\infty.
\een
Here $P_\mu$ may be identified with the ADM 4-momentum
and $U_{ij}$ and $V_{ij}$ are central charges which depend on the 
magnetic and electric charges $(p,q)$ and the moduli,
i.e. of the values $ {\bar{ \phi } _\infty }$ of the scalar 
fields at infinity. In the case of $N=2$ there are just two central charges
which may be combined into a single complex central charge
$Z(p,q,{\bar{ \phi _\infty }})$ and the Bogomol'nyi bound becomes
\ben
M \ge |Z(p,q,\bar{ \phi }_\infty )|. 
\een

\section{Finding Solutions}

We now look for local solutions on ${\cal M}= \Sigma \times {\Bbb R}$
which are independent of the
time coordinate $t \in {\Bbb R} $.
Globally of course 
the geometry is much more subtle because
of the presence of horizons but that will not affect the local equations
of motion.
The basic idea used here is that in 3-dimensions 
we may use duality 
transformations to replace vectors by scalars. The
resulting equations may be derived form an action describing
three-dimensional gravity on $\Sigma$ coupled to a sigma model.
 
\subsection{Reduction from 4 to 3 dimensions}

The metric is expressed as
\ben
ds^2 =-e^{2U} (dt + \omega _i dx^i)^2 + e^{-2U} \gamma_{ij} dx^i dx^j.
\een

The effective lagrangian is  
\ben
{ 1\over 16 \pi } R[\gamma] - {1\over 8 \pi }  
G _{ab}({\bar \phi}^a)  \partial _i 
{ \phi} ^b \partial_j { \phi} \gamma ^{ij},
\een
where ${ \phi}^a $ is the collection of fields 
$(U,\psi, {\bar \phi}^a, \psi^A, \chi _A)$ 
taking values in an augmented  target space $ M_{\phi}$
where $\psi$ is the twist potential, $\psi^A$ the 
electrostatic potential and  $\chi_A$ the magnetostatic potential.
The twist potential arises by dualizing $\omega_i$. 
Note that
$(U,\psi)$ are the gravitational analogues of electric and 
magnetic potentials
respectively. Indeed for pure gravity,
\ben
{\rm curl} \thinspace \omega = e^{-2U} {\rm grad} \thinspace \psi,
\label{twist} 
\een
the internal space is $ H^2 \equiv SO(2,1) /SO(2)$ and
the  metric is
\ben
dU^2 + e^{-2U} d \psi ^2\label{H2} .
\een
The  formula for the twist potential
becomes more complicated in the presence of vectors.

Three important general features to note are 

\begin{itemize}

\item If the signature of spacetime is $(3,1)$
then the signature of the $\sigma$-model metric $ G_{ab}$
is $(2+n_s, 2n_v)$. Physically this is because the $n_s$ scalars fields
$\phi^a$ and the gravitational scalars $(U, \psi)$ 
give rise to {\sl attractive} forces while the $n_v$ vector fields
give rise to {\sl repulsive} forces.

\item The metric ${\bar G}_{ab}$ admits $2 n_v +1$ commuting Killing fields
$ {\partial \over \partial \psi},{\partial \over \partial \psi^A},{\partial \over \partial \chi_A} $ which give rise to $2n_v+1$ charges,
the last of which, the so-called NUT charge, vanishes for asymptotically flat solutions.

\item Typically $ {M}_{\phi}$ is also a symmetric space
with indefinite metric
of the form ${ G} / { H}$, where $ G$ is an example  
U-duality group which includes both $S$ and $T$ duality groups.
Of course  $H$ is no longer  {\sl compact}.

\item If we were to include fermions we would get three-dimensional SUGRA
theory  
but with euclidean signature.

\end{itemize}

\subsubsection {Examples}

Let us look at some  examples of ${ G} /{ H}$.

\begin{itemize}

\item $N=1$. This is pure gravity,
there are no vectors and the signature is positive. The
the coset is two-dimensional hyperbolic space$H^2 =
SO(2,1)/SO(2)$.

\item $N=2$, $SU(2,1)/S(SU(1,1) \times U(1))$,   
Einstein-Maxwell theory. In fact  $\{M_\phi, G_{ab}\}$ is an analytic
continuation of the Fubini-Study metric on ${\Bbb C} {\Bbb P}^2$
to another real section with Kleinian signature $(2,2)$.
\item $N=4$ SUGRA, $SO(8,2) /SO(2) \times SO(6,2)$.
\item $N=4$ SUGRA plus supermatter, $SO(8,8) /SO(2,6) \times 
SO(6,2)$. This is what you get if you
dimensionally reduce $N=1$ supergravity from ten-dimensions.
\item The reduction of the Heterotic theory in 10-dimensions
gives, $ 
SO(24,8) /SO(22,2)\times SO(2,6)$
\item$ N=8$ SUGRA, $E_{8(+8)} /SO^*(16)$. 

\end{itemize}

It is clear that the group $G$ may be used as
a solution generating group. Of course
some elements of $G$ may not take physically 
interesting solutions
to physically distinct or physically interesting
solutions. nevertheless one may anticipate that
black hole solutions will
fall into some sort of multiplets
of a suitable subgroup
of $G$ and indeed this turns out to be the case.

\subsubsection{Static truncations}

In what follows we shall mainly be concerned with non-rotating
holes and so we drop the twist potential and consider the 
{\it static truncation} with effective lagrangian
\ben
+{ 1\over 16} R[\gamma] + 
{1 \over 8 \pi }( \partial U )^2 + 
{ 1 \over  8 \pi } {\bar G} _{ab} \partial {\bar \phi} ^a \partial {\bar \phi} ^b
- { 1\over 8 \pi } e^{-2U} (\partial \psi^A, \partial \chi _A ) {\cal M}^{-1}
(\partial \psi^A, \partial \chi _A )^t 
\een
  
\subsubsection{Gravitational Instantons}

The methods we have just described may also be used to obtain solutions
of the Einstein equations with positive definite signature
admitting a circle action. All that is required 
to get the equations is a 
suitable analytic  continuation of the previous formulae.
This entails a chage in the groups and the symmetric spaces.
Thus in the case of  pure gravity case the metric is
\ben
ds^2 = e^{2U} ( d \tau + \omega _i d x^i ) ^2 + e^{-2U} \gamma _{ij} dx ^i dx ^j.
\een
The twist potential still satisfies \ref{twist} but the internal
space becomes $dS_2= SO(2,1)/SO(1,1)$ with metric
\ben
ds^2= dU^2 - e^{-2U} d \psi ^2 \label{DS2}.  
\een

\subsubsection{The equations of motion}
The scalar equation of motion requires that $ \phi$ gives a harmonic
map from $\Sigma$ to $M_\phi$.
\ben
\nabla ^2 \phi =0
\een
where the covariant derivative $\nabla$ contains a piece
corresponding to the pull-back under $\phi$ of the connection
$\Gamma _{bc}^a(\phi)$ on $M_\phi$, thus 
\ben
\nabla _i  \partial _j \phi ^a= \partial_i \partial _j
- \Gamma _{ij}(x)  ^k \partial _k \phi^a + 
\partial _i\phi^c \partial _j \phi ^b\Gamma ^a_{bc} (\phi).
\een

Variation with respect to the metric $\gamma_{ij}$ gives
an Einstein type equation:
\ben
R_{ij} = 2\partial_\i \phi ^a \partial _j\phi  ^b G_{ab}.
\een
If we pretend that we are thinking of Einstein's equations
in three dimensions then the the left hand side may be thought
of as $T_{ij} - \gamma _{ij} \gamma ^{mn}T_{mn} $ where
$T_{ij}$ is the stress tensor.

There are essentially three easy types of solutions of this 
system of equations which may be described using simple geometrical
techniques.
\begin{itemize}
\item Spherically symmetric solutions
\item Multi-centre (i.e. BPS) solutions
\item cosmic string solutions.
\end{itemize}
\subsection{Spherically symmetric solutions}

The idea is to reduce the problem to one involving geodesics in $M_{\phi}$.
A consistent ansatz for the metric is $\gamma_{ij}$ is
\ben
\gamma _{ij} dx^i dx^j= { c^4 d \tau^2 \over \sinh ^4 c \tau}
+ { c^2 \over \sinh ^2 c \tau } ( d \theta ^2 + \sin ^2 \theta d \phi ^2 ).
\een
The radial coordinate $\tau$ is in fact a harmonic function on $\Sigma$
with respect to the metric $\gamma_{ij}$. 
The range of 
$\tau$  is from the horizon at $-\infty $ to spatial infinity at $\tau=0$.
In these coordinates   the 
only non-vanishing component of the Ricci tensor 
of $\gamma_{ij}$ is the radial component and this has the constant value
$ 2c^2 $. 
For a regular solution the constant
$c$ is related to the temperature and entropy by
\ben
c=2 ST.
\een

Now it is a fact about harmonic maps that the composition 
of a harmonic map with a geodesic map
is harmonic. 
Thus to satisfy the scalar equations of motion, $\phi^a(\tau)$ 
must execute
geodesic motion in $\{ G_{ab}, M_{\phi} \}$ 
with $\tau$ serving as affine parameter along the 
geodesic.
The Einstein equation then fixes the value
of the constraint
\ben
G_{ab} { d \phi ^a \over d \tau} { d \phi ^a \over d \tau}=c^2.
\een

Because of electromagnetic  gauge-invariance there are $2n_v$ Noether
constants of the motion, i,e, the electric and magnetic charges:
\ben
\pmatrix{ p^A \cr q_A \cr} = {\cal M} ^{-1} 
\pmatrix { { d \chi_ A \over d \tau}   \cr { d \psi ^ A  \over d \tau} \cr } .
\een

The remaining equations follow from
the effective action
\ben
\bigl ( { dU \over d \tau}  \bigr ) ^2 + {\bar G} _{ab} {d {\bar \phi} ^a \over d \tau } {d {\bar \phi} ^b \over d \tau } + e^{2U} V({\bar \phi},p,q)
\een
and the constraint becomes
\ben
\bigl ( { dU \over d \tau}  \bigr ) ^2 + {\bar G} _{ab} {d {\bar \phi} ^a \over d \tau } {d {\bar \phi} ^b \over d \tau } - e^{2U} V({\bar \phi},p,q)
=(2ST)^2.
\een

Evaluating this at infinity we get
\ben
M^2 + {\bar G}_{ab} \Sigma ^a \Sigma ^b -V({\bar \phi} _\infty, p,q)= ( 2S T)^2.
\een

The extreme case corresponds to $c=0$.
The metric $\gamma_{ij}$ is now 
\ben
\gamma_{ij}dx^i dx^j = { d \tau ^2 \over \tau ^4} + { 1 \over \tau ^2 } 
( d \theta ^2 + \sin \theta d \phi ^2 ).
\een

\subsection{Toda and Liouville systems}

The method just outlined has the advantage that it makes clear how 
the duality group acts on the solutions. 

The procedure also clear why
the radial equations frequently give rise to ordinary differential 
equations of Toda type which are in principle
exactly integrable.
As we have seen, the problem of finding spherically
symmetric solutions reduces to finding geodesics in the symmetric space
$G/H$. We may think of this as 
solving a a Hamiltonian system on on the co-tangent space
$T^\star(M_{\phi})=T^\star(G/H)$. 
This symmetric space admits at at least $2n_v + 1$ and typically more 
commuting Killing vectors. These come from the fact that
one may add an arbitrary constant to  the twist
potential $\psi$ and the magnetic and electric potentials 
$(\chi_a, \psi ^A)$. In addition there may be further symmetries
arising from axion like fields. Let us suppose that {\it in toto }
there are $r$ such commuting Killing vectors. 
Ignoring any possible identifications
they will generate
the group ${\Bbb R}^r$. One may eliminate the 
$r$ commuting constants of the motion to obtain
a dynamical system on the quotient configuration space
$ M_{\phi} /{\Bbb R}^r= {\Bbb R}^r \backslash G/H$. 
From a Hamiltonian point of view one is of course
just performing a Marsden-Weinstein symplectic reduction.

Now it is known from the work
of Perelomov and Olshanetsky that Toda systems arise precisely in this 
way. Thus it is no surprize that one encounters them in finding solutions
of the Einstein equations depending on a single variable.
The simplest example is when $M_{\phi}= G/H=SO(2,1)/SO(1,1)=AdS_2 $
or it's  Riemannian  version $ SO(2,1)/SO(2)=H^2$.
The internal metric is  given by  \ref{DS2} or \ref{H2} respectively. 
If the  constant of the motion $q= e^{-2U} { d \psi \over d \tau}$
then the dynamical system has effective Lagrangian
\ben
{ 1\over 2}  \bigl ( {dU \over d \tau}  \bigr )^2 \pm  { 1\over 2}
 q^2 e^{2U}
\een
with constant of the motion
\ben
{ 1\over 2}  \bigl ({ dU \over d \tau }\bigr )^2 \mp  { 1\over 2}
 q^2 e^{2U} = {\rm constant},
\een
where the upper sign refers the $AdS_2$ case and the minus
to the $H^2$ case. The resulting dynamical system is of course
a rather  trivial Liouville system and  may be integrated
using elementary methods.

If the internal space decomposes into a product of such models,
then the integration is equally easy. In practice most of the 
exact solutions in the literature may be obtained in this way.
As I mentioned above, in principle, the general Toda system is exactly
integrable but in practice it seems to be rather cumbersome to 
carry out the integration explicitly.

\subsection{ Multi-centre solutions}
We make the ansatz that the metric $\gamma _{ij}$ is flat
and we may take $\Sigma$
to be euclidean three-space ${\Bbb E}^3$. 
\ben
\gamma_{ij} =\delta_{ij}.
\een
This means
that the coordinates $(t,{\bf x})$ are harmonic coordinates,
i.e we are using a  gauge in which $ \partial _\mu \frak{g}^{\mu \nu} =0.$
The vanishing of the Ricci tensor
then requires the vanishing stress tensor or local force balance condition
\ben
\partial _i \phi ^a \partial _j\phi ^b G_{ab}=0.
\een

We must also satisfy the harmonic condition. The following construction
will do the job.We start
with $k$ ordinary harmonic functions $H^r(x^i)$, $r=1,2,\dots, k$
on Euclidean space ${\Bbb E}^3$. These give a harmonic map
from $ H: {\Bbb E}^3 \rightarrow  {\Bbb E}^k$.

We next find a  
$k$-dimensional  {\sl totally geodesic totally lightlike
} submanifold of the target space $M_\phi$. This is
a map $f: {\Bbb E}^k \rightarrow M_\phi$
whose image $N$ is    
\begin{itemize} 
\item{ Totally null, i.e the induced metric
$G_{rs}=G_{ab} {\partial f^a \over \partial y^r}{\partial f^b \over \partial y^s}
$ vanishes.}
\item { totally geodesic,
which means that a (necessarilly lightlike) geodesic which is initially
tangent to $N$ remains tangent to $N$.} 
\end{itemize}
The simplest example would be a null geodesic for which $k=1$.
The parameters $y^r$ are affine parameters.
In the case of Einstein Maxwell theory we may take the 
so-called \lq$\alpha$\rq  or \lq $\beta $\rq  2-planes which play a role in 
twistor theory. The important point about totally geodesic maps
is that they are harmonic.

Given our maps $H^r(x^i)$ and $f^a(y^r) $ we compose them,
i.e. we set
\ben
\phi^a(x^i) = f^a( H^r(x^i)).
\een
The result is a harmonic map and we are done.

This simple and elegant technique is in 
principle all that is required to construct 
all BPS solutions of relevance to four-dimensions.
As we shall see, it frequently works in higher dimensions.
Of course to check that they are BPS 
one has to check for the existence of Killing spinors.
The technique also  makes transparently clear the action of the 
U-duality group.
Moreover, since we are dealing with symmetric spaces,
everything can, in principle be reduced to calculations
in the Lie algebra of $G$.

Consider the simplest case, the static truncation of
Einstein Maxwell theory.
The internal space is $SO(2,1)/SO(1,1) = AdS_2$ with metric
\ben
dU^2 - e^{-2U} d \psi ^2
\een
where $\psi$ is the electrostatic potential.
The null geodesics are given by
\ben
\psi = { 1\over H}
\een
\ben
e^U= { 1\over H}.
\een
We have recovered the Majumdar-Papapetrou 
solutions.
For dilaton gravity with dimensionless coupling constant
$a$ which has the matter action
\ben
-{ 1\over 8p } \partial \sigma ^2 - { 1\over 16 \pi} e^{-2a \sigma} F^2 _{\mu \nu}
\een
the internal space carries the metric  (in the electrostatic case)
\ben
dU^2 + d\sigma ^2 - e^{-2 (a \sigma +U)}  d \psi ^2.
\een
The null geodesics are 
\ben
\psi = { 1\over \sqrt {1+a^2} }  { 1 \over H} 
\een
\ben
e^U= { 1\over H ^{ 1\over 1+a^2} }
\een
\ben
e^{-a \sigma}= { a^2 \over H ^{ 1\over 1+a^2} } 
\een
There are four interesting case:
\begin{itemize}
\item{ $a=0$: this is Einstein-Maxwell theory}
\item{ $a={1 \over \sqrt{3}} $: this is 
what you get if you reduce Einstein-Maxwell theory from five to four spacetime dimensions}
\item{ $a=1$: this corresponds to the reduction
of string theory from  ten to four spacetime dimensions}
\item{ $a=\sqrt{3} $: Kaluza-Klein theory. These
solutions are S-dual to the taub-NUT solutions, 
i.e. to Kaluza-Klein monopoles}
\end{itemize}
Modulo duality transformations,
these solutions are all special cases of the solutions with
four $U(1)$ fields and three-scalar fields. 
In the case that two $U(1)$ fields are electrostatic and 
two are magnetostatic the internal space 
decomposes as the metric product of four copies of $SO(2,1)/SO(1,2)$
. The spacetime metrics are given by:
\ben
ds^2 = -( H_1H_2 H_3 H_4)^{ -{1\over 4}} dt^2  + 
( H_1H_2 H_3 H_4)^{ {1\over 4}} d{\bf x}^2.
\een
The totally null, totally geodesic submanifolds are such that
\ben
( \psi^1, \psi^3, \chi_2 , \chi_4) = 
({ 1\over H_1},{ 1\over H_3},{ 1\over H_2},{ 1\over H_4}).
\een
The entropy is given by
\ben
S= {\pi }  \sqrt{q_1q_3p^2 p^4}.
\een

Of course these solutions may be lifted to 
eleven dimensions for example
where they may be thought of as intersecting
five-branes:.

\subsection {Other Applications}

The harmonic function technique works in other than
four spacetime dimensions. We now give a few examples.

\subsubsection{The D-Instanton} Perhaps the  simplest application  of the  technique described above  yields  the D-Instantons
of ten-dimensional type IIB theory. These are Riemannian solutions.
If $\tau= a+i e^{-\Phi}$ where $a$ is the pseudoscalar and $\Phi$ the
dilaton are the only excited bosonic
fields  other than the metric then
the Lorentzian equations come from the $SL(2,{\Bbb R})$-invariant 
action
\ben
R-{ 1\over 2} (\partial \Phi)^2  - e^{ 2\Phi} (\partial  a)^2.
\een
Note that we are using Einstein conformal gauge.

For the instantons $a=i\alpha$ with 
$\alpha$ real and the equations come from the action
\ben
R-{ 1\over 2} (\partial \Phi)^2  +e^{ 2\Phi} (\partial  \alpha)^2.
\een
This is effectively the same as the previous cases.
One takes  the Einstein metric  to be flat and

\ben
\alpha+ {\rm constant} =e^\Phi =H
\een
where $H$ is a harmonic function on ${\Bbb E}^{10}$.

Weyl rescaling the metric to string gauge

\ben
ds^2 =e^{{ 1\over 2} \Phi}d{\bf x}^2= H^{ 1\over 2} d{\bf x}^2
\een
gives an Einstein-Rosen Bridge down which global 
Ramond$\otimes$Ramond charge can be carried away. Note that 
large distances correspond to moderate string coupling
$g=e^{\Phi}$  while near
the origin of ${\Bbb E}^{10}$ corresponds to strong coupling
and the supergravity approximation  cannot be trusted.
 
\subsubsection{NS$\otimes$NS Five brane
in ten-dimensions}
The ten-dimensional metric in string conformal
frame is
\ben
ds_S^2=-dt^2 + (dx_9)^2+ (dx_8)^2+ (dx_7)^2+(dx_6)^2+(dx_5)^2
+ e^{2 \Phi} g_{\mu \nu} d x^\mu d x ^\nu 
\een
where $g_{\mu \nu}$ is the four dimensional metric in Einstein gauge.
If $a$ is the 4-dimensional dual of the NS$\otimes$NS three-form field 
strength then the equations follow from the lagrangian
\ben
R- 2 (\partial \Phi)^2 +{ 1\over 2} e^{4 \Phi} (\partial a)^2.
\een
Again one picks the metric $g_{\mu \nu}$ to be flat
and
\ben a+ {\rm constant}= e^{2\Phi} = H
\een
where $H$ is now a harmonic function on ${\Bbb E}^4$.

One may now apply a duality transformation taking
one to the Ramond$\otimes$ Ramond five-brane. This leaves the 
Einstein metric invariant but takes 
$\Phi \rightarrow - \Phi$. The resulting metric in string 
conformal gauge is 
\ben
ds^2 _S = H^{ -{1\over 2}} ( -dt^2 + 
(dx_9)^2+ (dx_8)^2+ (dx_7)^2+(dx_6)^2+(dx_5)^2) + 
H^{ 1\over 2} d {\bf x}^2. 
\een

\subsection{Cosmic string Solutions}

The seven-brane of Type IIB theory is an example of
of how to construct
cosmic string like solutions. The main difference
in technique  with the former
case  is that
since the internal metric is positive definite
the 
the spatial metric can no longer be flat
and so we need
to solve for it explicitly. This is simple if the spatial metric is two
dimensional.

To get the seven brane,
we write the ten dimensional metric in Einstein gauge as
\ben
ds^2 = -dt^2 + +(dx_9)^2 +(dx_8)^2+(dx_7)^2+(dx_6)^2+(dx_5)^2+(dx_4)^2+(dx_4)^2 + e^{ \phi} dz d{\bar z}.
\een
The static equations arise from the  two-dimensional Euclidean
action
\ben
R-{ 1\over 2} {(\partial \tau_1)^2 +(\partial \tau_2)^2 \over \tau_2^2}
\een
where $\tau=\tau_1+ i\tau_2= a+ i e^{-\phi}$ gives a map into the
fundamental domain of the modular group $SL(2,{\Bbb Z})  \backslash SO(2, {\Bbb R}) /SO(2)$.
We may regard  the two-dimensional space sections
as a Kahler manifold and the harmonic map equations are thus
satisfied  by a {\sl holomorphic ansatz}  $\tau= \tau(z)$.
We must also satisfy the Einstein condition.
Using the formula for the Ricci scalar of the two-dimensional
metric and the holomorphic condition this 
reduces to the {\sl linear} Poisson equation
\ben
\partial {\bar \partial }( \phi - {\log} \tau_2 )=0.
\een

To get the {\it fundamental string}
 one chooses
\ben
\phi=\Phi
\een
\ben \tau \propto \log z.
\een

In four spacetime time dimensions the fundamental string
is \lq super-heavy \rq, it is not asymptotically conical
at infinity.

To get the {\it seven brane}, which does correspond to
a more conventional cosmic string,  one picks
\ben
j(\tau(z)) = f(z)= { p(z) \over q(z) }
\een
where $j(\tau)$ is the elliptic modular 
function and $ f(z)= { p(z) \over q(z)}$ is a 
rational function of degree $k$.

The appropriate solution for the metric is
\ben
e^{\phi} = \tau _2 \eta ^2{\bar \eta}^2 \Bigl |  \prod ^{i=k} (z-z_i) ^ 
{ -{1 \over 12} }
    \Bigl |^2.
\een
where $\eta (\tau)$ is the Dedekind function.
Asymptotically
\ben
e^\phi \sim (z {\bar z} ) ^ {-{ k \over 12}}.
\een
Therefore the  spatial metric is that of a cone with deficit angle
\ben
\delta = { 4k \pi \over 24}.
\een
This may also be verified using the equations of motion and the 
Gauss-Bonnet theorem.  As a result one  can have up to 
12 seven-branes in an open universe.
To close the universe one needs 24 seven-branes.

The solution has the following \lq F-theory \rq interpretation.
One considers the 
metric 
\ben
ds^2 = g_{ij} dy^i dy^j + e^{\phi} dz d{\bar z}, \label{stringy}
\een
where $g_{ij}$ is the following unimodular metric on
the torus $T^2$ with coordinates $y^i$
\ben
\pmatrix{
\tau_2^{-1}& \tau_1 \tau_2^{-1}\cr
\tau_1 \tau_2^{-1}& \tau_1^2 \tau _2 ^{-1} + \tau _2 \cr
}.
\een
The metric \ref{stringy}is self-dual or hyper-K\"ahler.
If one takes 24 seven-branes 
one gets an approximation to a K3 surface
elliptically fibered over ${\Bbb C } {\Bbb P}^1$.

Another interesting special case arises
as an orbifold. 
Consider $T^2 \times {\Bbb C}$ with coordinates
$(y^1, y^2, z)$.  Quotient by the involution 
$(y^1, y^2, z) \rightarrow (-y^1, -y^2, -z)$.
There are  four fixed points which may be blown up
to obtain a regular simply connected
manifold on which there exists a 
twelve real-dimensional
family of family
of smooth hyper-K\"ahler metrics. 
The second Betti number is five and the 
intersection form of the five non-trivial cycles
is given by the Cartan matrix
of the extended Dynkin diagram ${\bar {D_4}}$.
These metrics have been obtained as 
hyper-K\"ahler quotients by Kronheimer and by Nakajima. 
The smooth
metrics are rotationally symmetric
but $\partial \over \partial y^1$ and $\partial \over \partial y^2$
are only approximate Killing vectors. 
physically they are interesting as examples of \lq Alice Strings \rq
because, thinking of the two-torus as  Kaluza-Klein type internal space
with two approximate $U(1)$ 's and  two approximate
charge conjugation
operators $C_2 : (y^1, y^2, z) \rightarrow (-y^1, y^2, z)$ and 
$C_2  : (y^1, y^2, z) \rightarrow (y^1,- y^2, z)  $,
one finds that if Alice circum-ambulates  
the string, but staying very far away, she returns  
charge conjugated.  Of course if she ventures into the core region
she will find that the two electric charges are not strictly conserved.
Because
the solutions admit a non-triholomorphic circle action,
they are given in terms of a solution of the $su(\infty)$ Toda
equation. From this it is easy to check that
the solutions approach the orbifold limit with exponential
accuracy.

To see this explicitly note that the  metric
\ben
ds^2= { 1\over \nu ^\prime} (2 d \theta + \nu _1 dy^2 - \nu_2 dy^2 ) ^2
+ \nu ^\prime \Bigl \{ d \rho ^2 + e^\nu \bigl ( (dy^1)^2 + (d y^2)^2 \bigr ) \Bigr \}
\een
is hyperK\"ahler if $\nu( \rho, y^1,y^2)$ satisfies
\ben
\bigl ( e^\nu \bigr )^{\prime \prime } + \nu _{11}+  \nu _{22}=0 \label{toda},
\een
where the superscript $\prime$ denotes differentiation with 
respect to $\rho$ and the subscripts $1$ and $2$ denote 
differentiation with 
respect to $y^1$ and $y^2$ respectively. 
The Killing vector
$\partial \over \partial \theta$ leaves invariant
the  privileged  K\"ahler form
\ben
(2 d \theta + \nu_1 d y^2 - \nu _2 dy^1 ) \wedge  d \rho + 
+\bigl ( e^ \nu \bigl ) ^\prime d y^1 \wedge d y^2.
\een
whose closure 
requires that \ref{toda} is true. Note that the geometrical
significance of the coordinate $\rho$ is that it
is the moment map for the circle action.
The simplest solution of (\ref{toda}) is $e^\nu =\rho$.
this gives the flat metric
\ben
ds^2= dr^2 + r^2 d\theta^2 + (dy^1)^2 +(dy^2)^2, 
\een
with $r=2 \sqrt{\rho}$. This is independent of $y^1$ and $y^2$.
For solutions admitting an elliptic fibration 
we require a solution of (\ref{toda}) which is periodic in  $y^1$ and $y^2$.
To check the typical behaviour near infinity,
one linearizes (\ref{toda}) about the solution $e^\nu =\rho$.
The resulting equation admits 
solutions by separation of variables. It is then a routine exercise
to convince one's self that the general solution must decay
exponentially at infinity.

Before  leaving these metrics it is perhaps worth
pointing out that their relation to the much better known
class of Ricci-flat
riemannnian metrics admitting a tri-holomorphic circle action  
and which  depend on an arbitrary
harmonic function $H$ on ${\Bbb E}^3$. They are easily obtained using
the technique described above.  The metrics are
\ben
ds ^2 = H^{-1} ( d t +\omega_i d x ^i)^2 + H {\bf dx} ^2 \label{harmonic},
\een
with
\ben
{\rm curl } \thinspace \omega = {\rm grad }\thinspace  H \label{curl}.
\een
with Kahler forms
\ben
H d x^1 \wedge d x^2 + dx^3 \wedge (dt + \omega_i d x^i).
\een
\ben
H d x^2 \wedge d x^2 + dx^1 \wedge (dt + \omega_i d x^i).
\een
\ben
H d x^3 \wedge d x^1 + dx^2 \wedge (dt + \omega_i d x^i).
\een
The closure of these K\"ahler-forms is equivalent to
the condition \ref{curl}.

If $H$ is independent of $\arctan ( x^2/y^2)$
there  will be an additional  circle action which
preserves  the first K\"ahler form 
but rotates the second into the third.
This means that by taking an arbitrary
axisymmetric harmonic function we can, in principle,
obtain a solution of the the $su(\infty)$ toda equation \ref{toda}.

To get a complete metric one must
choose $H$ to be a finite sum
of $k$ poles with identical positive residue.
The coordinate singularities at the poles may then be removed by 
periodically identifying the imaginary time coordinate $t$.  
If $H \rightarrow 1$ at infinity the metrics are asymptotically 
locally flat, \lq ALF \rq. and represent $k$
Kaluza-Klein monopoles. If $ H \rightarrow 0 $ at infinity
the metrics are asymptotically locally euclidean, \lq ALE \rq.
Thus if $0 \le t \le 2 \pi$ and $h= { 1\over 2r}$ 
we get the flat metric on ${\Bbb R}^3$
while $H= 1 + { 1\over 2r} $ we get
the Taub-NUT metric on ${\Bbb R}^3$.

On the other hand if we take $\phi= \log \tau_2$ and 
identify $z=x^1+i x^2$ and $ t=y^1$ and $y^2= x^3$ 
the metrics \ref{stringy} amd \ref{harmonic}
 coincide. In fact  $H=\tau _2$ and 
$\omega _3= \tau_1$.

We see that $\partial \over \partial y^1 $
generates a triholomorphic circle action.
The three K\"ahler forms are
\ben
\Omega _1 =
\tau_2 d x^1 \wedge d x^2 + dy^2 \wedge dy^1,
\een
\ben
\Omega_2= \tau_2 dx^2 \wedge dy^2 + d x^1 \wedge (d y^1 + \tau _1 d y^2)  
\een
and 
\ben
\Omega _3=\tau_2 dy^2 \wedge dx^1 + d x^2 \wedge (d y^1 + \tau _1 d y^2).  
\een
and they are closed by virtue of the Cauchy-Riemann equations for
$\tau(x^1 +i x^2)$.

\section {Conclusion}

In these two lectures I have tried to give some idea of 
what qualifies as a soliton in 
classical  supergravity theories and how one finds the solutions. I have
concentrated on general principles and 
largely restricted myself to four spacetime dimensions. 
The lectures were emphatically {\sl not}
 not intended as a comprehensive review. For
recent applications the reader is referred to other articles
in this volume or to the voluminous current literature.
Appended below is a rather restricted list of references 
largely confined to papers that I have written, 
either alone or with collaborators, where the reader may find more details
of the claims made above or from which the reader may trace
back to the original sources.  As I stated above, it was not my intention
to provide a comprehensive review and no slight is intended
against those not explicitly cited.


\end{document}